\immediate\write18{makeindex \jobname.nlo -s nomencl.ist -o \jobname.nls} 

 \documentclass[final,3p,times,twocolumn]{elsarticle}

\usepackage{amssymb}
\usepackage[fleqn]{amsmath}
\usepackage{epstopdf}
\usepackage{graphicx}
\usepackage{caption}
\usepackage{subcaption}
\usepackage{float,lscape}
 \usepackage{gensymb}
\usepackage{amsmath}
\usepackage{soul}
\usepackage{color, soul}
\usepackage{siunitx}

    \usepackage{framed} 
    \usepackage{multicol} 

    \usepackage{nomencl} 

    \makenomenclature
    \renewcommand*\nompreamble{\begin{multicols}{2}}
    \renewcommand*\nompostamble{\end{multicols}}
    \usepackage{ifthen}

	\renewcommand{\nomgroup}[1]{
           \ifthenelse{\equal{#1}{S}}{\item[\emph{Subscripts}]}{
           \ifthenelse{\equal{#1}{G}}{\item[\emph{Greek letters}]}{                
          \ifthenelse{\equal{#1}{U}}{\item[\emph{Superscripts}]}}}}

\makeatletter
\newcommand*{\rom}[1]{\expandafter\@slowromancap\romannumeral #1@}
\makeatother

\journal{}

\begin{document}

\begin{frontmatter}

\title{Contaminant transport by human passage through an air curtain separating two sections of a corridor: Part II -  two zones at different temperatures}


\author[label1,label2]{Narsing~K.~Jha\corref{cor1}}
\ead{navinnaru88@gmail.com}
\author[label1]{\rm D.~Frank}
\author[label3]{L.~Darracq}
\author[label2]{P.~F.~Linden}

\address[label1]{~Department~of~Applied~Mathematics ~and~ Theoretical ~Physics,~ University ~of ~Cambridge,~ Wilberforce ~Road,~ CB3~ 0WA ~Cambridge,~ UK}
\address[label2]{~Present~Address:~Department ~of ~ Physics~of ~ Complex~systems,~Weizmann~Institute~of~Science,~ Rehovot,~Israel}
\address[label3]{Department ~of ~Mechanical ~Engineering, ~Ecole ~Polytechnique, ~France}


\begin{abstract}

Air curtains are installed in open doorways of a building to reduce buoyancy-driven exchange flows across the doorway. Although an air curtain allows an unhampered passage of humans and vehicles, the interaction of this traffic with an air curtain is not well understood. We study this problem by conducting small-scale waterbath experiments with fresh water and salt water solutions. As a model of human passage, a vertical cylinder is pulled through a planar jet representing an air curtain and separating two zones at different densities. For a fixed travel distance of the cylinder before and after the air curtain, the average infiltration flux of dense fluid in light fluid side increases with increasing cylinder velocity. However, we find that the infiltration flux is independent of density difference across the doorway and the travel direction of the cylinder. As a consequence, the sealing effectiveness of an air curtain reduces with an increasing cylinder speed and this reduction is independent of the direction of the buoyancy-driven flow. Dye visualisations of the air curtain and the cylinder wake are used to examine the re-establishment process of the air curtain after its disruption by the cylinder. We observe that the re-establishment time of the air curtain and the infiltration in the cylinder wake increases with an increasing cylinder speed.

\end{abstract}


\end{frontmatter}


\section{Introduction}

Buoyancy-driven exchange flows across an open doorway between two zones at different temperatures result in unnecessary heat losses and transport of particulate substances, which can have detrimental effects in many settings such as in industrial manufacturing processes as well as in laboratory or hospital clean rooms. Air curtains are commonly used to minimise these buoyancy-driven exchange flows and to reduce the spread of unwanted agents. A major advantage of an curtain over mechanical installations, such as vestibules, sliding doors or strip curtains, is that it allows an unhindered passage of human and vehicular traffic through the doorway. However, the interaction of the air curtain with a moving object such as a person walking across the doorway and through the air curtain is not well understood. In particular, a moving object is accompanied by a wake that entrains and mixes ambient fluid and thus contributes to the contaminant transport across the doorway. In the companion paper ``Part I - uniform ambient temperature” \citep{Jha2020part1}, we considered the situation when a person walks through the doorway between two sections of a corridor at the same temperature and, therefore, at the same density, and thus transports contaminants between two zones in their wake. We studied how the air curtain can be used in that scenario to reduce the contaminant transport. In this present second part of our study, we investigate the situation when a density difference is additionally present across the doorway driving a buoyancy-driven exchange flow and, as before, a person walks across the doorway and through the air curtain. 

The simplest configuration of an air curtain is a downward blown planar air jet which is produced by a centrifugal fan mounted in a manifold above the doorway. Fundamental features of turbulent jets are well understood \citep{rajaratnam1976}, however their usage as an air curtain is still under investigation because of the further complication of wall impingement \citep{khayrullina2017piv} and different densities on either side of the jet. According to \citet{HayStoe1,HayStoe2}, the key parameter to decide the performance and stability of an air curtain is the deflection modulus $D_m$  defined as the ratio of the jet momentum flux and the pressure difference due to buoyancy difference across the doorway,

\begin{equation}
\begin{split}
D_m=\frac{\rho_0 b_0 u_0^2}{g H^2 \left(\rho_d-\rho_l\right)}=\frac{(\rho_0 {Q}_0^2/b_0)}{g H^2 \left(\rho_d-\rho_l\right)}
\\=\frac{{Q}_0^2}{g b_0 H^2 \left(\frac{T_0}{T_d}-\frac{T_0}{T_l}\right)}.
\label{eq:Dmdef}
\end{split}
\end{equation}
 Here, $u_0$, $\rho_0$, $T_0$ and $Q_0$ are the discharge velocity, density, temperature (in Kelvin) and volumetric discharge per unit nozzle length of the air curtain jet at the manifold exit, respectively. The nozzle width is denoted by $b_0$, the door height by $H$ and $g$ is the acceleration due to the gravity. For a doorway between two zones at different temperatures the subscripts $d$ and $l$ are used to denote the properties of the dense (cold) and light (warm) air, respectively. From the value of $D_m$, it can be determined whether the air curtain is stable (for $D_m\gtrapprox0.15$) or not (for $D_m\lessapprox0.15$), i.e. whether the air curtain reaches the floor and impinges on it \citep{FrankLinden}. 

The performance of an air curtain is usually quantified by the sealing effectiveness $E$, which is defined as the ratio of the fraction of the flow prevented by the air curtain compared to the flow through an unprotected doorway,
\begin{equation}
E = \frac{q - q_{ac}}{q},
\label{eq:Edef}
\end{equation}
where $q_{ac}$ and $q$ are the exchange flows through the doorway with and without the air curtain, respectively. The effectiveness $E$ first increases with $D_m$ due to the disruption of the organised buoyancy-driven flow by the curtain until a maximum value of $E$ is reached. With a further increase in $D_m$, $E$ decreases. Under optimal operating conditions, an air curtain can reduce the buoyancy-driven exchange flow by about 80$\%$ compared to an open doorway situation.

Examining the flow at high $D_m$, \citet{GuyonnaudSolliec} highlighted the importance of shear layer eddies in the jet and the jet impingement on the floor on the effectiveness $E$. \citet{SirenPartI,SirenPartII} presented  methods for dimensioning an air curtain using momentum and moment-of-momentum balance principles. He also determined the minimum momentum required for the air curtain to reach the opposite side of the doorway, both in the presence and absence of a wind. \citet{SirenPartII} focused on the thermal behavior of the air curtain deriving the expression for the thermal loss and comparing it with empirical results. Full-scale experiments have been performed by \citet{HowellShibata} and \citet{FosterOptimum}, which showed satisfactory agreement with the theoretical predictions of \citet{HayStoe1}. Numerical simulations \citep{CostaOliveira,foster2007cfd,Goncalves}, a semi-analytical model  \citep{GiraldezSemiAnalytical} and laboratory-scale studies in water \citep{FrankLinden,frank2015effects} are all in satisfactory agreement with full-scale tests on an air curtain in an otherwise sealed building. 

\citet{FrankLinden} studied the effect of an additional ventilation pathway, such as an open window, on the effectiveness of an air curtain and modelled the observed change of performance of the air curtain caused by the resulting change in the neutral level. Further, \citet{frank2015effects} investigated a heated air curtain and observed the reduced stability, effectiveness and energy efficiency of the curtain associated with the opposing buoyancy force. Additionally, wind and pressurised chambers provide external forcing to the air curtain and can destabilise it causing severe oscillations \citep{havet2003experimental,rouaud2006behavior}. \citet{Qi2018} studied the effect of a human presence below an air curtain (but not the the process of a person walking through the air curtain) and they observed that the person had either no influence or reduced the infiltration due to the additional blockage of the doorway. The interaction between the wake of a moving person and an air curtain installed in the doorway separating two zones at different densities has not been studied and is the motivation for the present work. 

We conducted laboratory waterbath experiments to investigate the interaction of a moving person, modelled as a cylinder, with the air curtain, that separates two zones at different densities in a corridor. The paper is structured as follows. In section \ref{S:2}, we describe the experiments and the techniques used for flow visualization and measurements of the contaminant transport. The experimental results are presented in section \ref{S:3}. In section \ref{sec:Vi}, we quantify the infiltration volume $V_i$ from one section to the other when a person moves through the air curtain in the doorway along a fixed travel distance. We examine different travel velocities, density differences across the doorway and directions of travel. In section \ref{sec:Edef} we discuss the correct definition of the air curtain effectiveness in the present scenario and, subsequently, examine how effective an air curtain is in section \ref{sec:Effectiveness}. In section \ref{sec:ComparisonIandII}  we briefly establish a connection between the experiments reported in Part I \citep{Jha2020part1} and the present experiments. Dye visualizations of the air curtain jet and the infiltration wake are described in section \ref{sec:Dye}. Finally, in section \ref{S:4} we summarise our conclusions, provide an overview of how the present study can be helpful for a better design of an air curtain, and discuss strategies for the reduction of the traffic effects on the air curtain effectiveness in practical conditions. 

\section{Experimental methods}
\label{S:2}

Experimental measurements and methods are explained in detail in the companion paper hereafter referred to as Part I \citep{Jha2020part1}. We briefly summarise here the main components of the experimental setup and some additional features of the present set of experiments in which there is a horizontal density difference across the doorway.

\begin{figure*}
\centering
\includegraphics[scale=0.4]{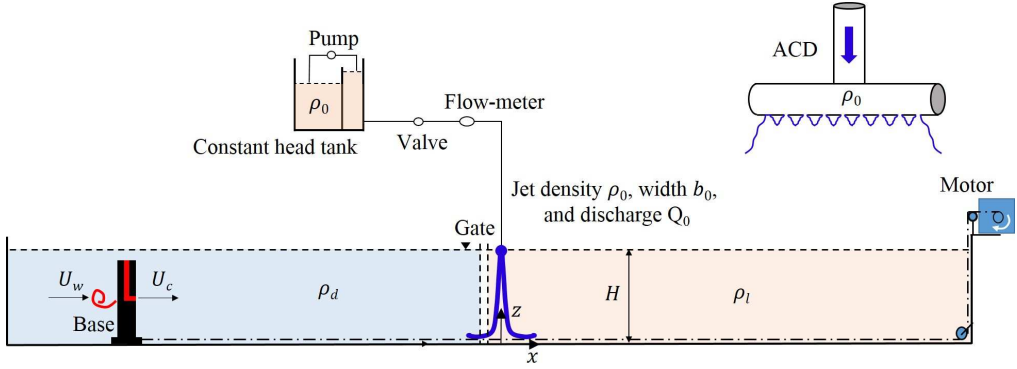}
\caption{Schematic showing the experimental setup. Experiments were conducted in a long rectangular channel, each half of which was filled with dense and light fluid of density $\rho_d$ and $\rho_l$, respectively. Fluid of density $\rho_0$ was supplied to the air curtain device (ACD) from a constant head tank. A vertical gate was installed beside the curtain to separate the fluid when the air curtain device was switched off. The cylinder was driven by a motor, connected by a flexible thread and a system of pulleys. A dye port injecting red dye was attached to the cylinder to track the wake and the infiltration. Blue coloured dye was injected at the nozzle exit to visualise the air curtain.}
\label{fig:ExpSetup}
\end{figure*}

A schematic of the experimental setup is shown in figure \ref{fig:ExpSetup}: a long rectangular channel (of width $W$ and length $2L$) was divided in two equal sections by a vertical gate, which represented a doorway in a corridor. One side of the tank was filled with fresh water of density $\rho_l$ and the other side with salt water of density $\rho_d>\rho_l$, creating a horizontal density difference $\Delta\rho=\rho_d-\rho_l$. The density ratio $\rho_l/\rho_d$ was kept in the range $0.96 - 1$ to maintain the validity of the Boussinesq approximation.

Details of the cylinder passage to represent a moving person, the dye visualisation methods and the air curtain device (ACD) are described in Part I. For the present set of experiments, the ACD was supplied with fresh water of density $\rho_0$, so that here $\rho_0=\rho_l<\rho_d$. The cylinder was pulled between two fixed points (the midpoints of both sides) both from the salt water side to the fresh water side of the tank and vice versa. This arrangement models the situation of a person entering or leaving a warm building through a warm air curtain from or to a cold outside environment. Table \ref{tab1} summarises parameters that we used for the present set of experiments.

\begin{table*}[!ht]
	\begin{center}

		\begin{tabular}{lccccccccc}
		\hline
		Series & $\rho_d$ & $\rho_l$& $\rho_0$& $Q_0$ & $D_m$ & $t_{fw}$& $w_m=H/t_{fw}$ & $U_c$ &$U^*$ \\
		
		& $\SI{}{\kilo\gram\per\meter\cubed}$ & $\SI{}{\kilo\gram\per\meter\cubed}$ & $\SI{}{\kilo\gram\per\meter\cubed}$ & $\SI{}{\milli\meter\squared\per\second}$ & &$\SI{}{\second}$ &  $\SI{}{\milli\meter\per\second}$ &  $\SI{}{\milli\meter\per\second}$ &  \\
		[3pt]
		\hline
		
		Series A & 1002 -- 1041 & 998 & 998 & 425 & 0.05-1.2 & 0.525 & 400 & $0$ & $0$ \\ 
		
		Series B & 1002 -- 1041 & 998 & 998 & 425 & 0.05-1.2 & 0.525 & 400 & $89$ & 0.22 \\ 
		
		Series C & 1002 -- 1041 & 998 & 998 & 425 & 0.05-1.2 & 0.525 & 400 & 135 &  0.34 \\ 
		
		Series D & 1002 -- 1041 & 998 & 998 & 425 & 0.05-1.2 & 0.525 &  400 & $174$ &  0.43 \\ 
		
		\hline
		\end{tabular}
		\caption{The experimental conditions and parameter values used in experiments. Here, $t_{fw}$ is the time taken by the curtain to reach the bottom of the tank $z/H=0$ from the cylinder height $z=l$ in a quiescent environment of the same density, $H$ is the height of the curtain from base of the tank, $\rho_l$ and $\rho_d$ are the densities of the fluid in the dense and light side of the tank, respectively.}
		\label{tab1}
	\end{center}
\end{table*}

		
%
%

The experimental procedure was similar to experiments E6 -- E15 in Part I. One side of the tank was initially filled with salt water of density $\rho_d$ up to the height of $H=\SI{210}{\milli\meter}$ and the other side with fresh water of density $\rho_l$ to a slightly lower level of  $\SI{205}{\milli\meter}$. The ACD was located in the fresh water side was switched on and, once the jet flow was steady and the water level in both sides equilibrated, the vertical gate was opened. The blue dye port to track the air curtain was switched on and the time measurement started. For Series A (see table \ref{tab1}) no cylinder motion took place, for Series B, C, D, the cylinder was set into motion at a constant velocity $U_c$, and the red dye was injected to visualise the cylinder wake. For Series A, we finished the experiment after approximately $t=\SI{10}{\second}$. For the other Series, the experiment ended when the cylinder reached the midpoint of the opposite side after passing through the air curtain. The vertical gate was then closed, and the air curtain as well as both dye ports were switched off. The total duration time $t$ of an experiments (in the order of magnitude of $\SI{10}{\second}$) was measured using a stop watch to a precision of $\SI{0.1}{\second}$. After the gate was closed, the water in both sections was mixed and the new densities $\rho_{ln}$ and $\rho_{dn}$ were then measured. 

We investigated both directions of travel for the cylinder: from the dense, (equivalently cooler) salt water side ($\rho_d$) to the lighter, (equivalently warmer) fresh water side ($\rho_l$) and vice versa. We varied the value of the deflection modulus $D_m$ by changing the density $\rho_d$ from $\SI{1002} - \SI{1041}{\kilogram\per\cubic\meter}$.

Using the mass conservation for the light fluid side (considering $\rho_{0}$ = $\rho_{l}$) of the tank and assuming zero net flow across the doorway, the infiltration volume $V_i$ of dense fluid into the light fluid side during an experimental run was calculated from the measured initial and final densities as

\begin{equation}
    {V_i = \left(V_l +\beta q_0t\right)\frac{(\rho_{ln} -\rho_{l})}{(\rho_{d} -\rho_{l})}.}
    \label{eq:Vi-A}
\end{equation}
Here, $q_0$ is the total source volume flux (in $\SI{}{\milli\meter\cubed\per\second}$) of the air curtain and $V_l$ is the initial volume of the light fluid half of the tank. As explained in Part I, $\beta$ denotes the fraction of the air curtain volume flux that spilled into the light fluid compartment and was set to $\beta=0.5$ without introducing a measurement error of more than 3\%.

As in Part I, we calculate the average exchange flux during an experimental run as
\begin{equation}
    {q_{ac}=\frac{V_i}{t},}
\end{equation}
where $V_i$ is given in (\ref{eq:Vi-A}) and $t$ is the measured duration of the experiment. We recall from Part I that the the total infiltration flux $q_{ac}$ can be decomposed as

\begin{equation}\label{eq:QacDecomp}
    {q_{ac}=q_{ac,m}+q_{ac,cyl},}
\end{equation}
where $q_{ac,m}$ is the exchange flux caused by the mixing process within the air curtain and $q_{ac,cyl}$ is the infiltration flux due to the interaction of the air curtain with the wake of the cylinder. In absence of a moving cylinder in Series A, $q_{ac}=q_{ac,m}$.

As explained in Part I the experimental parameters were chosen to provide dynamical similarity with a person walking through a doorway at full scale. We recall from Part I that we use the non-dimensional cylinder speed $U^* \equiv U_c/w_m$. The mean velocity of the air curtain front is $w_m \equiv H/t_{fw}$, where $t_{fw}$ is the time needed by an establishing air curtain to reach the tank bottom $z/H=0$ from the cylinder top height $z=l$, with $z$ being the vertical coordinate. We varied the cylinder speed from $\SI{89}{\milli\meter\per\second}$ to $\SI{174}{\milli\meter\per\second}$, which results in values of $U^*$ from 0.22 to 0.43. At full scale, walking at a moderate pace of $\SI{1}{\meter\per\second} = \SI{3.6}{\kilo\meter\per\hour}$ through a typical air curtain with the speed $w_m \sim \SI{2}{\meter\per\second}$, gives $U^* \sim 0.5$. 


\section{Results}
\label{S:3}

\subsection{Infiltration volume and flux measurements}\label{sec:Vi}

\begin{figure*}[ht]


\centering
\includegraphics[scale=0.6]{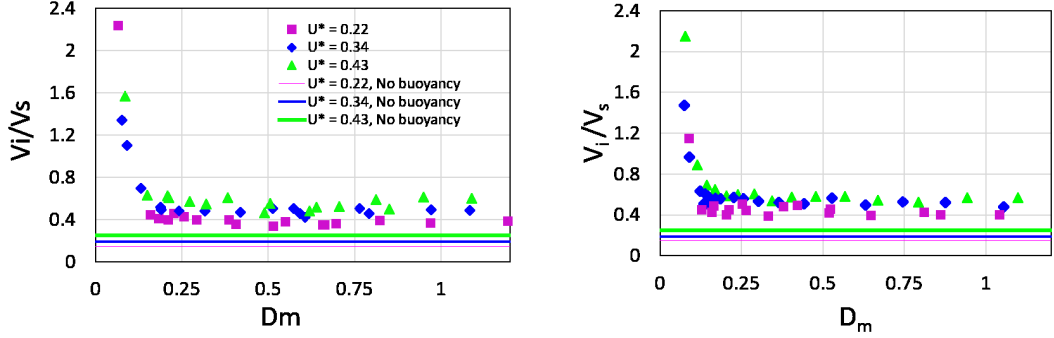}
\caption{The non-dimensionalised infiltration volume and infiltration flux $V_i/V_s=V_i/(ldU_c t)=q_{ac}/(ldU_c)$ for an operating air curtain for different cylinder speeds $U^*$ as a function of the non-dimensional horizontal density difference. The direction of the cylinder motion is in (a) from $\rho_d$ to $\rho_l$ and in (b) vice versa. The data for the cases of no horizontal difference (solid lines) have been interpolated from the experimental data measured in Part I.}
\label{fig:Var_q_ac}
\end{figure*}

Figure \ref{fig:Var_q_ac} shows the measured infiltration volume $V_i$ for experiments of Series B, Series C and Series D. Figure \ref{fig:Var_q_ac}a plots the data for the travel direction of the cylinder from the dense fluid half to the light fluid half and figure \ref{fig:Var_q_ac}b for the opposite direction against $D_m$. Since we change $D_m$ by varying $\rho_d$, the data are effectively plotted against a (reciprocal value of a) varying horizontal density difference $\Delta\rho$. As in Part I, we non-dimensionalise $V_i$ by the swept volume of the cylinder $V_c=ldL=ldU_ct$, where $l$ is the cylinder height, $d$ its diameter and $L$ the travel distance or half the total length of the experimental tank. Note that
\begin{equation}
    \frac{V_i}{V_s}=\frac{V_i}{ldU_ct}=\frac{q_{ac}}{ldU_c}.
\end{equation}

 We observe that once the air curtain is stable (for $D_m \gtrapprox 0.2$), the infiltration volume $V_i/V_s=q_{ac}/(ldU_c)$ rises slightly with $U^*$ (and hence $U_c$ for a fixed air curtain strength). This is similar to our observation in Part I that for low cylinder speeds $U^*$ the infiltration volume increases slightly with the cylinder speed. Furthermore, $V_i/V_s=q_{ac}/(ldU_c)$ does not depend on the horizontal density stratification across the air curtain and remains constant for varying $D_m$. Moreover, the travel direction of the cylinder also appears to be irrelevant and we measure same values of $V_i/V_s=q_{ac}/(ldU_c)$ for both travel directions. As a consequence of  $V_i/V_s=q_{ac}/(ldU_c)$ being independent of $D_m$, and since
\begin{equation}
    q_{ac}=q_{ac,m} +q_{ac,cyl},
\end{equation}
we note that $q_{ac,cyl}$ due to the cylinder wake is also independent of the horizontal density stratification. This is because once the air curtain is stable, the flux due to the mixing by the air curtain $q_{ac,m}$ is independent of the density difference.

The infiltration volume $V_i/V_s=q_{ac}/(ldU_c)$ being independent of $D_m$ and the travel direction of the cylinder is a remarkable observation. We explain this phenomenon by noting that although the moving cylinder disrupts the air curtain, the wake behind the cylinder is still a turbulent flow field. We expect this turbulence in the cylinder wake to inhibit and destroy the directed buoyancy-driven current -- similar to how the air curtain itself prevents the buoyancy-driven flow in the absence of a moving cylinder. We will confirm this later using dye visualisation (figure~\ref{fig:DyeDm}), where we observed a noticeable effect of the cylinder passage but little evidence of any buoyancy-driven exchange flow.

\subsection{Air curtain effectiveness}\label{sec:Edef}

The air curtain effectiveness is defined as the fraction of the flow prevented by the air curtain compared to the open-door situation (\ref{eq:Edef}),
where $q$ is the buoyancy-driven flux. Conventionally, $q$ is calculated using the theoretical expression, also known as the orifice equation

\begin{equation}
    {q = \frac{1}{3}C_{d}A\sqrt{g{'}H},}
    \label{eq:orifice}
\end{equation}
where $A$ is the area and $H$ is the height of the opening, respectively. The experimentally measured discharge coefficient is $C_d\approx 0.6$ for sharp-edged openings \citep{Linden1999fluid}.

In Part I we studied the situation when the air curtain is installed between two zones with the ambient fluid at the same density and, in the absence of an air curtain, the exchange flow is solely due to the wake of a moving person, modelled as a moving cylinder in our small-scale experiments. There, we defined the air curtain effectiveness as

\begin{equation}
    E_c=\frac{q_{cyl}-q_{ac}}{q_{cyl}}=1-\frac{q_{ac}}{q_{cyl}},
    \label{eq:EcDef}
\end{equation}
where $q_{cyl}$ is the exchange flux associated with the wake of the moving cylinder.

In our present experiments we have both contributions to the exchange in the absence of an air curtain: buoyancy-driven flow due to the horizontal density across the doorway and the exchange flow due to the wake of a moving cylinder. Hence, the question arises whether (\ref{eq:Edef}) or (\ref{eq:EcDef}) would be more appropriate to calculate the air curtain effectiveness in this combined situation.

\begin{figure}[!ht]
\centering\includegraphics[width=0.925\linewidth]{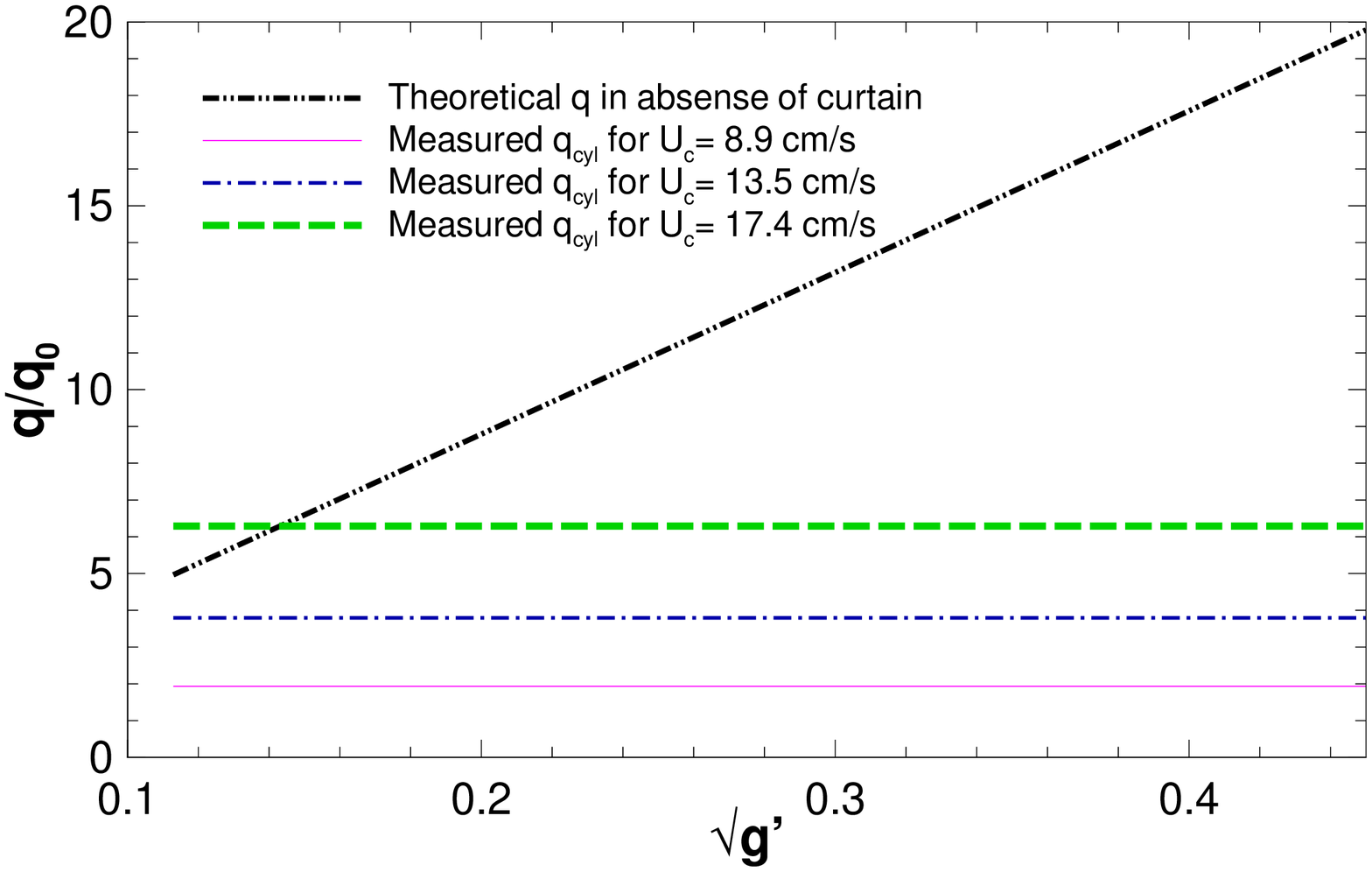}\\
\caption{Comparison of the buoyancy-driven exchange flow rate $q$ for the range of $\Delta\rho$ used in our experiments and the infiltration flow rate $q_{cyl}$ due to the moving cylinder for the range of $U^*$ in our experiments. The theoretical line for $q$ was calculated using (\ref{eq:orifice}). The horizontal lines show the experimentally measured $q_{cyl}$ values for $\Delta\rho=0$ in Part I. We expand these values as lines across a range of $g'$ for an easier visual comparison between $q_{cyl}$ and $q$. {$q_0$ is the volumetric flow rate of the curtain.}}
\label{fig:QvsQcyl}
\end{figure}

Figure \ref{fig:QvsQcyl} shows the comparison between the theoretically calculated buoyancy-driven exchange flux using (\ref{eq:orifice}) for the range of $\Delta \rho$ in our experiments and the measured $q_{cyl}$ (see Part I) for different $U_c$ for the unprotected doorway. For the present opening configuration, we used $C_d \approx 0.55 \pm0.03$ in the orifice equation (\ref{eq:orifice}), see \ref{app:DischCoeff}. In figure \ref{fig:QvsQcyl}, we observe that for our set of parameters, the infiltration flux due to the density difference in the absence of an air curtain dominates the infiltration flux due to the cylinder wake. Thus, we conclude that the air curtain effectiveness should be based on the buoyancy-driven exchange flux and the formula (\ref{eq:Edef}) should be used.

\subsection{Effectiveness measurements}\label{sec:Effectiveness}

\begin{figure}[!ht]
\centering
\begin{subfigure}[]{0.5\textwidth}
\centering
\includegraphics[scale=0.5]{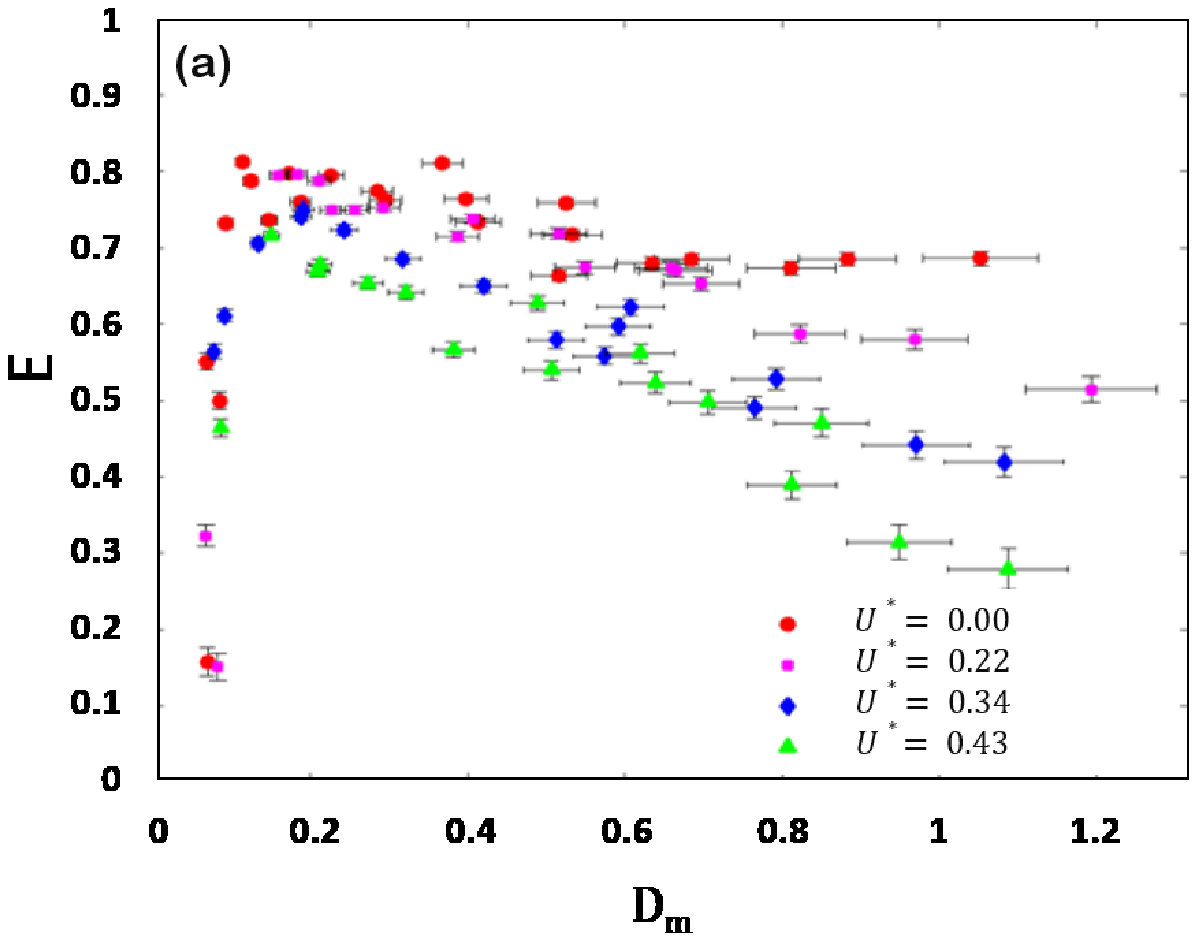}
\caption{}
\label{fig:EDm-a}
\end{subfigure}

\begin{subfigure}[]{0.5\textwidth}
\centering
\includegraphics[scale=0.5]{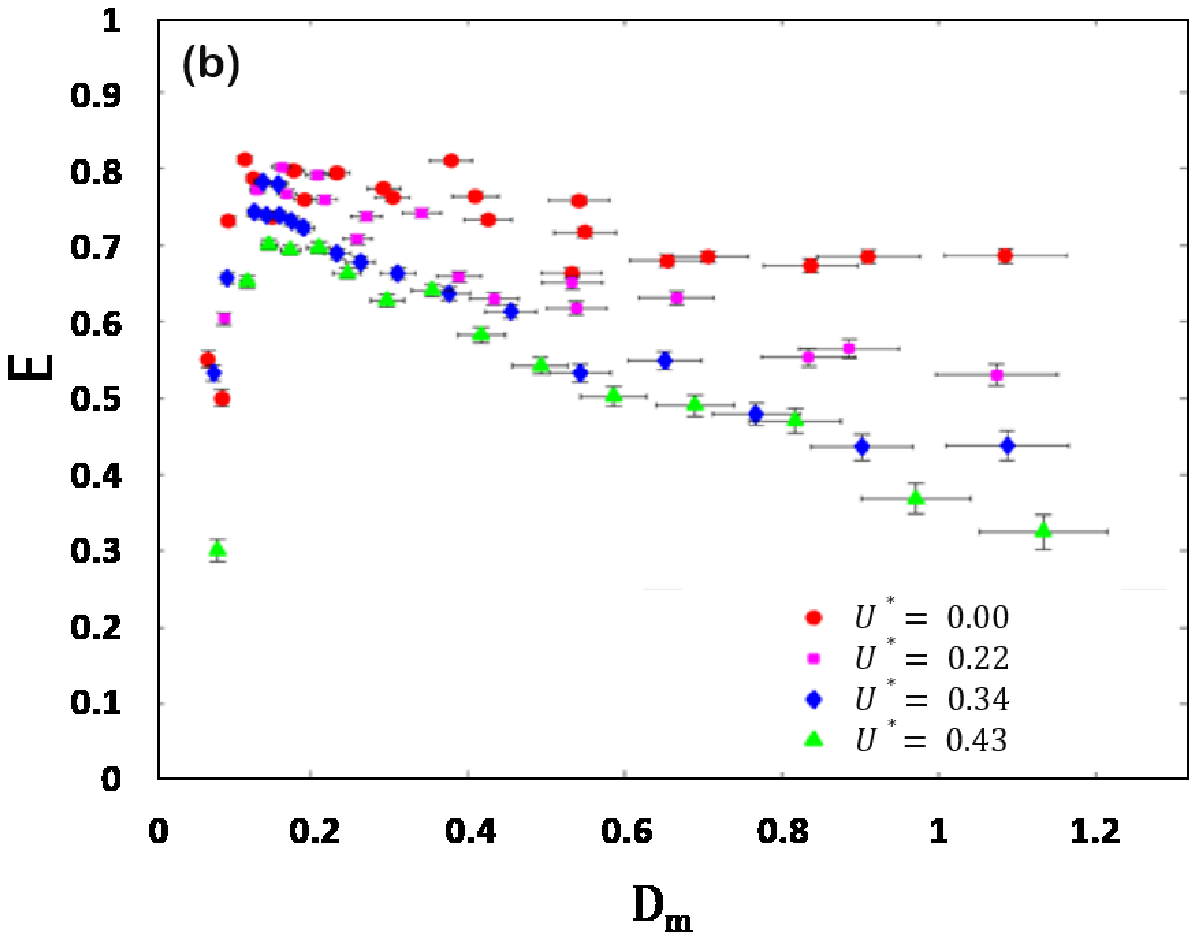}
\caption{}
\label{fig:EDm-b}
\end{subfigure}

\caption{ Air curtain effectiveness $E$ as a function of the deflection modulus $D_m$ for different dimensionless cylinder speeds $U^* \equiv U_c/w_m $: (\ref{fig:EDm-a}) the cylinder moves from the dense fluid to the light fluid side while in (\ref{fig:EDm-b}) the motion is in the opposite direction. In these experiments the values of the deflection modulus $D_m$ were changed by changing the density difference across the curtain -- see (\ref{eq:Dmdef}). Consequently, the error bars reflect the {uncertainty} in the density difference during the course of an experiment.}
\label{fig:EDm}
\end{figure}

The variation of the air curtain effectiveness $E$ with the deflection modulus for different cylinder speeds is shown in figures \ref{fig:EDm-a} and \ref{fig:EDm-b} for the transits from the dense to the light fluid side and vice versa, respectively. Recalling (\ref{eq:Edef}), the effectiveness is
\begin{equation}
    E=1-\frac{q_{ac}}{q}=1-\frac{\left(q_{ac,m}+q_{ac,cyl}\right)}{q},
    \label{eq:Edecomp}
\end{equation}
where we use the decomposition (\ref{eq:QacDecomp}). 

For case of the air curtain operating without the cylinder transit in Series A, $q_{ac,cyl}=0$, the effectiveness $E$ first increases with the deflection modulus until it reaches a maximum value of about 0.8 at $D_m \sim 0.2$, and then decreases slowly with further increase in $D_m$. This value of maximum effectiveness and the dependence on $D_m$ is similar to the values measured previously at a doorway between two rooms by \citet{FrankLinden} and others. This similarity between the present case of a doorway in corridor and doorways between (wider) rooms suggests that the exchange across the doorway is not sensitive to the room configurations and is determined locally at the doorway itself. At the maximum effectiveness, the curtain is stable and impinges on the floor. As mentioned previously, with further increase in $D_m$, the effectiveness $E$ reduces because of the enhanced mixing between two compartments due to the jet entrainment and impingement at the bottom, i.e., an increasing $q_{ac,m}$. 

The dependence of the effectiveness $E$ on $D_m$ is shown in figure \ref{fig:EDm} for different values of the dimensionless cylinder speeds $U^* \equiv U_c/w_m$. There is no observable difference between the air curtain effectiveness in figure \ref{fig:EDm-a} and in figure \ref{fig:EDm-b} for opposite directions of the cylinder travel. This is the reflection of the fact that the infiltration volume $V_i$  and the infiltration flux $q_{ac}$ do not depend on the cylinder travel direction as was discussed and explained in section \ref{sec:Vi} and figure \ref{fig:Var_q_ac}.

In general, the effectiveness $E$ was reduced by the passage of the cylinder as a result of the increased transport across the doorway in the cylinder wake, $q_{ac,cyl}>0$, and, for a given value of $D_m$, the reduction increased with increasing cylinder speed (recall from figure \ref{fig:Var_q_ac} that $q_{ac}/(ldU_c)$ rises slightly when the air curtain is stable and, hence, $q_{ac}$ increases with $U_c$). At small values of the deflection modulus, $D_m\lessapprox 0.15$, the air curtain is inherently unstable and, thus, the motion of the cylinder had no noticeable effect on the effectiveness. For deflection modulus $D_m \approx 0.2$  where the effectiveness is maximum, $E$ was reduced by approximately 10\% at the highest cylinder velocity. At higher values of the deflection modulus the reduction in $E$ increased with increasing $D_m$. At $D_m>1$ the effectiveness was reduced by about 25\% compared to the base case for $U^*=0.22$, and more than 55\% for $U^*=0.43$.

Figures \ref{fig:Var_q_ac} and \ref{fig:QvsQcyl} reveal why the influence of a moving cylinder is more noticeable for high $D_m$ values in figure \ref{fig:EDm}. In our experiments, we achieve $D_m\to\infty$ by $\Delta\rho\to 0$.  In that limit, $q$ becomes small (figure \ref{fig:QvsQcyl}) but $q_{ac}$ remains constant (figure \ref{fig:Var_q_ac}). Invoking (\ref{eq:Edecomp}), the contribution $q_{ac}/q$ increases for an increasing $D_m$ (by reducing $\Delta\rho$), and thus $E$ displays a significant decrease. This provides a simple explanation of the trend of the $E(D_m)$ curves in figure \ref{fig:EDm} specifically for our experimental arrangement in which we increase $D_m$ by reducing $\Delta\rho$. In the following, we discuss a more general argument for the case when $D_m$ is increased by, for example, increasing the curtain momentum flux $b_0 u_0^2$.

We shall assume that the geometric configuration of the doorway and the cylinder is fixed, i.e., the door height $H$, the door width $W$, the doorway area $A=WH$,  the cylinder diameter $d$, the cylinder height $l$ and the air curtain nozzle width $b_0$, which is a reasonable assumption for any real-case scenario. We also consider only the regime in which the air curtain stably impinges on the bottom of the tank. Using (\ref{eq:Dmdef}), (\ref{eq:orifice}), and the definition of the modified deflection modulus for the cylinder motion introduced in Part I as
\begin{equation}
    {D_{m,c}=\frac{b_0 u_0^2}{U_c^2 H}\times\frac{H}{l}\times\frac{W}{d}=\frac{b_0 u_0^2W}{U_c^2 ld}}
    \label{eq:Dmc_def}
\end{equation}
we can re-write the effectiveness $E$ as

\begin{align}
      {E} &{= 1 - \frac{q_{ac}}{q}}\nonumber\\
      &{= 1 - \frac{3}{C_d}\cdot\frac{q_{ac}}{W\sqrt{H}\sqrt{g'H^2}}}\nonumber\\
      &{= 1 - \frac{3}{C_d}\cdot\frac{q_{ac}}{W\sqrt{H}\sqrt{b_0 u_0^2}}\sqrt{D_m}}\nonumber \\
      &{= 1 - \frac{3}{C_d}\cdot\frac{q_{ac}\sqrt{W}}{W\sqrt{H}\sqrt{ld U_c^2}}\frac{\sqrt{D_m}}{\sqrt{D_{m,c}}}}\nonumber\\
      &{= 1 - \Gamma\cdot\frac{q_{ac}}{AU_c}\frac{\sqrt{D_m}}{\sqrt{D_{m,c}}}},
\end{align}
where $\Gamma = 3/C_d\cdot\sqrt{WH}/\sqrt{ld}$ includes the parameters for the fixed doorway and cylinder geometry. By further observing

\begin{equation}
   U^* = \frac{U_c}{w_m}\sim\frac{U_c}{u_0}\sim\frac{1}{\sqrt{D_{m,c}}},
\end{equation}
so $U*=C_1/\sqrt{D_m}$ for some constant $C_1$, we can deduce
\begin{align}
    {E} & {= 1 - C_2\cdot\frac{q_{ac}}{U_c}\sqrt{D_m}U^*}\nonumber\\
      & {\approx 1 - C_2 C\cdot\sqrt{D_m}U^*,}
      \label{eq:EDm_th}
\end{align}
where $C_2=C_1\Gamma$. By going to the last line in (\ref{eq:EDm_th}) we make an empirical argument that  $q_{ac}/U_c\approx C$ is not expected to undergo significant variations for the range of $U_c$ of practical interest. This was demonstrated experimentally in figures \ref{fig:Var_q_ac}.

Taking the derivative for a constant $\sqrt{D_m}$
\begin{equation}
    {\frac{\partial E}{\partial U^*}\bigg|_{\sqrt{D_m}}\approx - C_2 C\cdot\sqrt{D_m}}
\end{equation}
shows that the decrease in $E$ as the consequence of a moving cylinder with speed $U^*$ is more pronounced for large values of $D_m$. This explains the observed changes in $E$ in figure \ref{fig:EDm}.

\subsection{Infiltration flux $q_{ac,cyl}$ due to the cylinder wake and relationship between the experiments in Part I and Part II}\label{sec:ComparisonIandII}

In section \ref{sec:Vi} we noted that $q_{ac}$ is independent of the  density difference across the air curtain. In particular, this implies that also $q_{ac,cyl}$ is expected to be independent of $\Delta\rho$. In this section we show how $q_{ac,cyl}$ can be extracted from our experiments and establish a connection between the experiments in Part I and the experiments presented here in Part II.

\begin{figure}[!ht]
	\centering\includegraphics[scale=0.375]{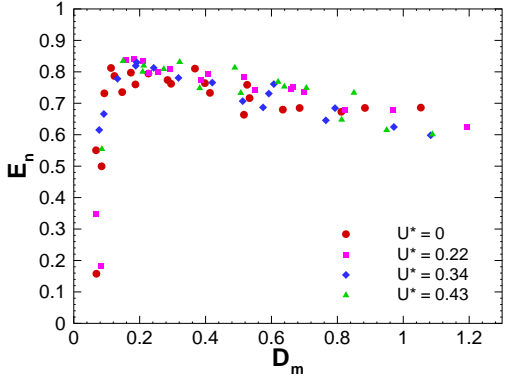}
		\caption{ Corrected effectiveness curves $E_n(D_m)$ of an air curtain for different dimensionless cylinder speeds $U^*$ as calculated using (\ref{eq:En}). The corrected effectiveness curves collapse onto the air curtain effectiveness curve for $U^*=0$, i.e., the case with no moving cylinder.}
	\label{fig:EnDm}
\end{figure}

Following the procedure for experiments in Part I ($\Delta\rho=0$), we conduct one experiment with just an operating air curtain ($Q_0=\SI{480}{\milli\meter\squared\per\second}$) and no moving cylinder, $U^*=0$. In this particular experiment, we have $q_{ac,m}=q_{ac}$ since the whole exchange flux between two halves of the tank is only due to the mixing by the air curtain. We interpolate the measured data for $q_{ac}$ for $Q_0=\SI{480}{\milli\meter\squared\per\second}$ in experiments in Part I (experiments E11, E12, E13, E14, E15) to obtain the interpolated values for $q_{ac}$ for $U^*=0.22,\textrm{ }0.34,\textrm{ }0.44$ ($Q_0=\SI{338}{\milli\meter\squared\per\second}$). We can now estimate
\begin{equation}
    {q_{ac,cyl}\big|_{U^*>0}\approx q_{ac}\big|_{U^*>0}-q_{ac,m}\big|_{U^*=0}.}
\end{equation}
{Subsequently, we use these $q_{ac,cyl}$ values to calculate the corrected effectiveness values for experiments here in Part II for $U^*=0.22,\textrm{ }0.34,\textrm{ }0.44$ as}
\begin{equation}
    {E_n= 1-\frac{q_{ac,m}+q_{ac,cyl}}{q}+\frac{q_{ac,cyl}}{q}=E+\frac{q_{ac,cyl}}{q} ,}
    \label{eq:En}
\end{equation}
{where $E$ are the effectiveness values presented in figure \ref{fig:EDm-a}.}
The corrected data $E_n(D_m)$ are presented in figure \ref{fig:EnDm}. We recognise that when the effectiveness $E$ corrected by $q_{ac,cyl}$ (constant for each $U^*$), all the $E_n(D_m)$ curves collapse within the experimental error onto the effectiveness curve $E(D_m)$ for $U^*=0$ (i.e., when there is no moving cylinder). This confirms that there is no strong nonlinear interaction between the wake and the gravity current.

\subsection{Dye visualisation and the air curtain re-establishment process}\label{sec:Dye}

\begin{figure*}[!ht]
\centering
\includegraphics[scale=0.55]{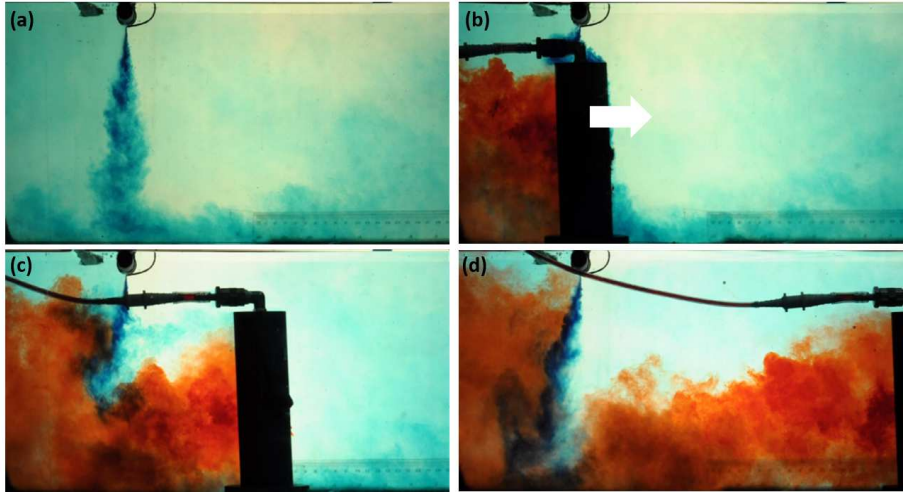}
\caption{Side views of the interaction between the cylinder and the curtain for $U^*= 0.44$ and $D_m = 0.5$. The cylinder is moving from the dense to the light fluid side (from left to right as marked in (b)). The blue-dyed curtain jet separates the dense and the light fluid in (a) and the red dye visualises the cylinder wake and the infiltration. The infiltration across the curtain and the re-establishment process is shown in (c). The curtain is re-established in (d).}
\label{fig:DyeSide}
\end{figure*}

\begin{figure*}[!ht]
\centering\includegraphics[width=0.95\linewidth]{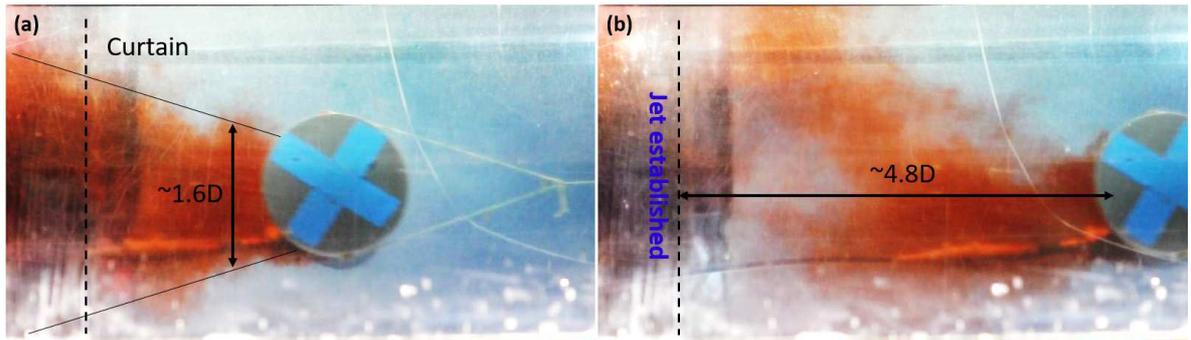}
\caption{Plan views of the interaction between the cylinder and the air curtain for the same parameters as in figure~\ref{fig:DyeSide}, $U^*=0.44$ and $D_m=0.5$. The wake expands away from the cylinder and thus the spanwise infiltration across the curtain increases with time.  The curtain is re-established in (b). The marked dimensions are in cylinder diameters.}
\label{fig:DyePlan}
\end{figure*}

\begin{figure*}[!ht]
\centering
\includegraphics[scale=0.5, trim={0.14cm 5.2cm 0 0},clip]{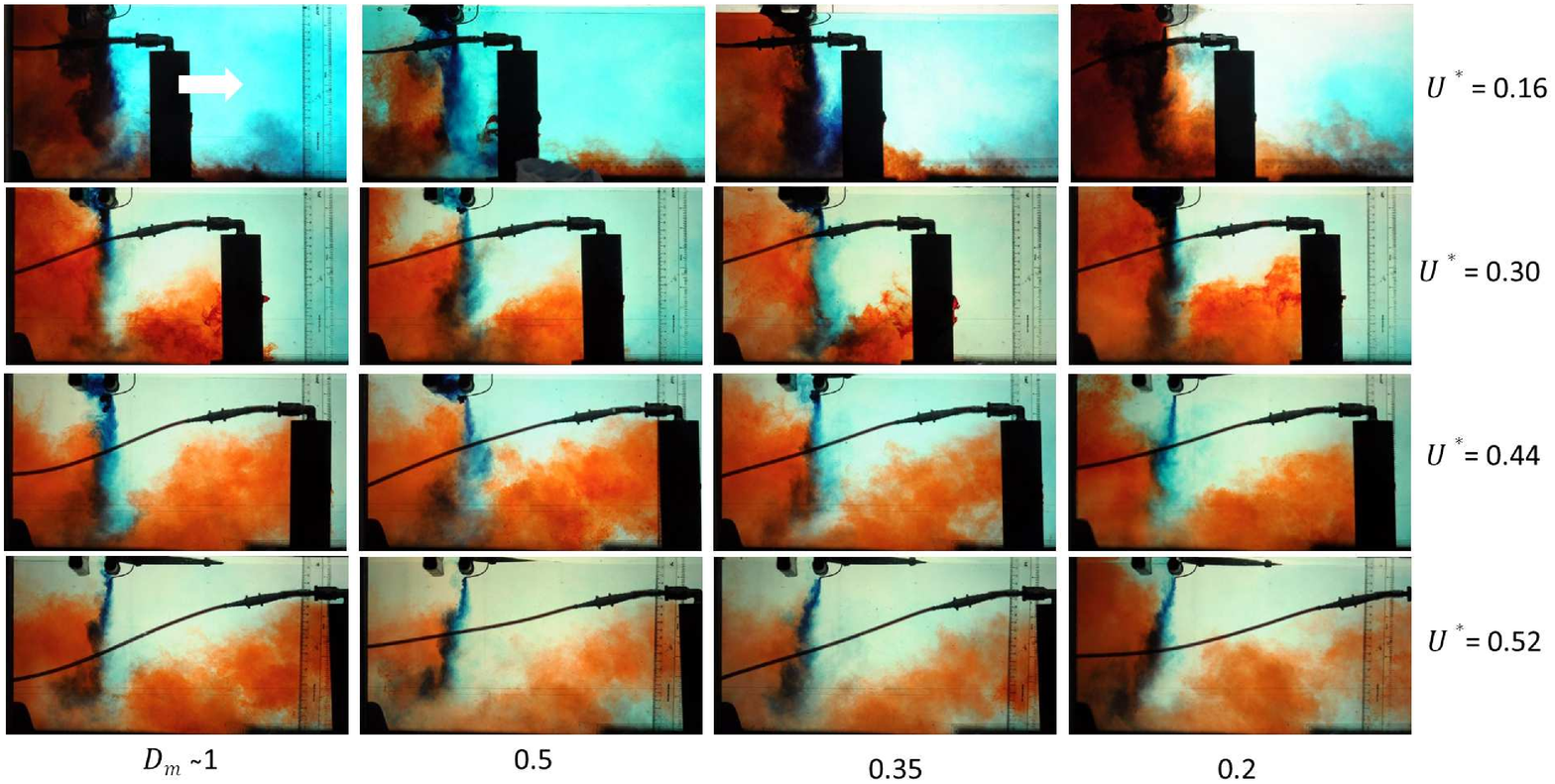}

\centering
\includegraphics[scale=0.5, trim={0.14cm 0 0 15.25cm},clip]{Dye_comp_established.eps}
\caption{Side views of the cylinder wake and the resulting infiltration when the curtain is first re-established for a range of cylinder speeds and deflection moduli. The cylinder is moving from the dense to the light fluid side and from left to right in the figure as marked by the white arrow.}
\label{fig:DyeDm}
\end{figure*}

We now present the dye visualisations and use them to gain further insight into the flux and effectiveness measurements. Flow visualisation was conducted for three different cylinder speeds and for four different deflection modulus values. The infiltration flux $q_{ac}$ and the infiltration volume $V_i$ do not depend significantly on the horizontal density difference $\Delta\rho$ (figure \ref{fig:Var_q_ac}). Likewise, the direction of travel does not seem to affect the infiltration flux (figures \ref{fig:Var_q_ac} and \ref{fig:EDm}), so we only present the results for the cylinder moving from the dense to the light fluid side of the tank. 

Side views of the passage of the cylinder (left to right, marked by the arrow in figure~\ref{fig:DyeSide}(b)) are shown in figure \ref{fig:DyeSide}. The cylinder was set into motion  $\SI{0.5}{\meter}$ away from the air curtain which was not influenced by the cylinder at that distance as can be seen in figure \ref{fig:DyeSide}(a). As the cylinder approached the air curtain, there was a small deflection of the air curtain in the direction of motion of the cylinder. However, the major disruption occurred as the cylinder passed below the air curtain as shown in figure \ref{fig:DyeSide}(b). Immediately after the cylinder passed through the air curtain, the infiltration of dense (red) fluid into the light fluid side took place during the period of re-establishment of the curtain.  At later times, the air curtain re-established by penetrating the wake as seen in figure \ref{fig:DyeSide}(c). Figure~\ref{fig:DyeSide}(d) illustrates the moment when the air curtain was first re-established. A colour movie \emph{`Movie1'} showing the complete infiltration process can be found in the supplementary material. 

Plan views of the interaction for the same parameter values are shown in figure~\ref{fig:DyePlan}. As was explained in Part I, the cylinder possesses a base with a larger diameter, which obscures the view immediately adjacent to the cylinder. The wake width increases with the distance from the cylinder, so that the span-wise width of the infiltration increased with time. In this case the air curtain is re-established (figure~\ref{fig:DyePlan}(b)) after the cylinder has traveled about 5 cylinder diameters beyond the air curtain. 

\begin{figure*}[!ht]
  \centering
  \begin{subfigure}[t]{\linewidth}
        \centering\includegraphics[width=0.7\linewidth]{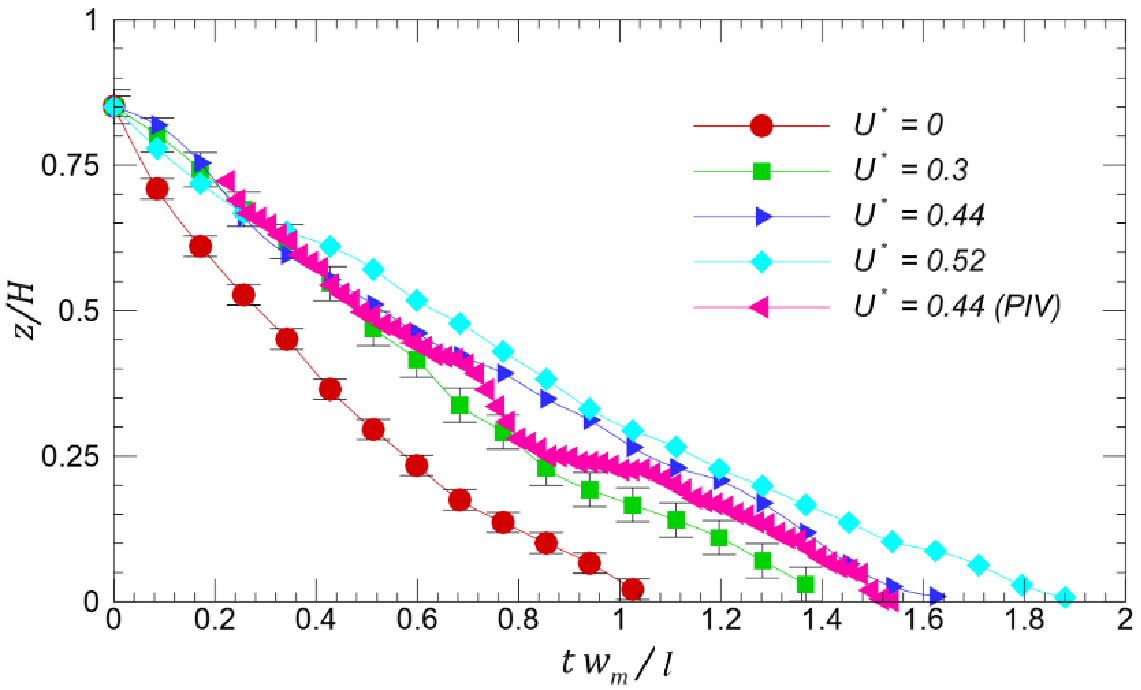}
	\caption{ Re-establishment of the curtain as a function of time after the cylinder passage. The front of the curtain is tracked and the time $t=0$ corresponds to the time when the jet front is at the height of the top of the cylinder, which is $0.15H$ from the nozzle exit. The evolution of the jet is shown for $D_m=0.5$, $q_0 = \SI{425}{\milli\meter\tothe{2}\per\second}$  and for different cylinder speeds $U^*$ = 0.3, 0.44, 0.52. }
\label{fig:Re-establishment-a}
        \end{subfigure}%
        \vspace{4mm}
  \begin{subfigure}{\linewidth}
           \centering\includegraphics[scale=0.95, trim={0.04cm 0 0 0.04cm},clip]{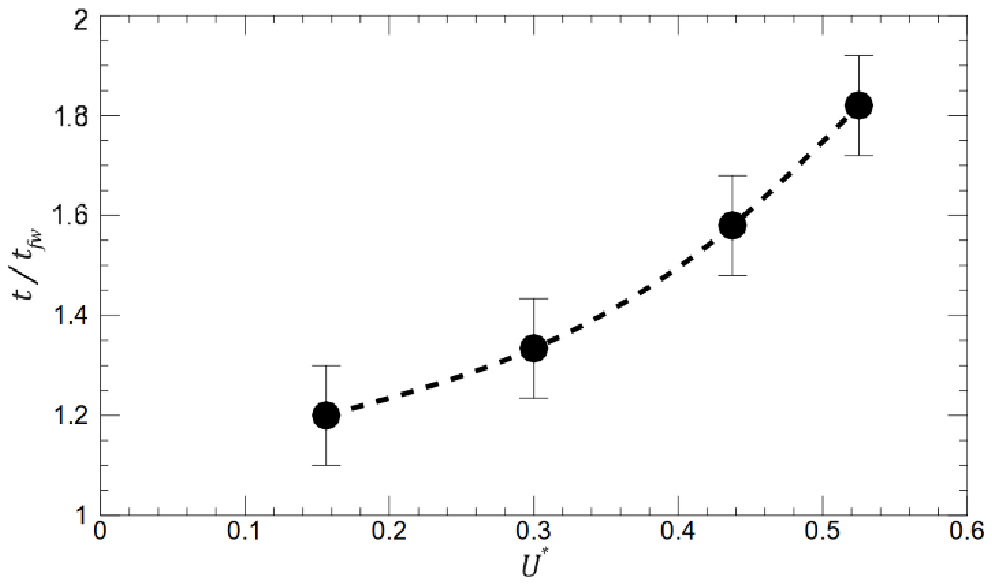}
            \caption{Time taken for the curtain to reach the bottom of the tank for the same experimental parameters as in figure \ref{fig:Re-establishment-a}. Here, the time is non-dimensionalised by the time $t_{fw}$ taken for the curtain to reach the tank bottom from the cylinder head in a quiescent fresh water environment.}
\label{fig:Re-establishment-b}
        \end{subfigure}
     \vspace{4mm}
\caption{The curtain re-establishment characteristics for different cylinder speeds. Lines are shown to describe the trends of the data and also to demarcate among different conditions. }
\label{fig:Re-establishment}
\end{figure*}

Figure \ref{fig:DyeDm} shows the disruption and the re-establishment of the air curtain for different cylinder speeds and values of the deflection modulus. These snapshots show the infiltration when the curtain is first re-established. For a given value of the deflection modulus $D_m$, the amount of the red dyed fluid on the right hand side of the air curtain increases with increasing cylinder speed. This is in line with the experimentally observed slight increase in the infiltration volume $V_i/V_s$ in figure \ref{fig:Var_q_ac} for a fixed $\Delta\rho$ (corresponding to a fixed $D_m$). However, there is no noticeable influence of the deflection modulus value, which in the present case is varied by changing $\Delta\rho$, for a fixed $U^*$. This is to be expected since as shown in figure \ref{fig:Var_q_ac}, the infiltration volume flux $V_i/V_s=q_{ac}/(ldU_c)$ does not vary significantly with $g'$, and hence $\Delta\rho$, for a range of values. We stress, however, that a constant $q_{ac}$ (with $\Delta\rho$) does not imply a constant $E$, so $E$ will still vary with $D_m$.

The re-establishment of the air curtain after its disruption determines the total amount of the infiltration by the wake. The dimensionless vertical penetration height $z/H$ of the jet, measured upwards from the tank bottom $z/H=0$ so that the jet nozzle is at $z/H=1$, is plotted as a function of time for different cylinder speeds in figure~\ref{fig:Re-establishment-a}. The measurement for the case $U^*=0$ was carried out in the fresh water environment without any density difference. This is also the measurement from which we determine $t_{fw}$. For the other cylinder speeds, we fixed $D_m=0.5$. Here, we also briefly present the air curtain re-establishment time for $U^*=0.44$ from the PIV measurements reported in Part I to compare it with the front tracking using dye visualisation in figure~\ref{fig:Re-establishment-a}. 
We recall that the PIV experiment was conducted in the absence of any buoyancy effects for $\Delta\rho=0$. For PIV, we track the curtain front at 100 f.p.s., taking every fourth frame from the PIV recordings, whereas the front tracking from the dye visualisation is done at 24 f.p.s. In the PIV experiment, the visible bump in the measured curve is due to the presence of a strong eddy near the curtain front. Apart from that, it is very similar to the dye tracking, which includes the presence of buoyancy. This again confirms that the presence of a horizontal density difference $\Delta\rho$ does not significantly affect the curtain interaction with the cylinder wake. The total time required for the curtain to re-establish is plotted in figure~\ref{fig:Re-establishment-b}. The increasing re-establishment time of the air curtain with the increasing cylinder speed is in line with the measured decrease in its effectiveness.

\section{Summary and conclusions}\label{S:4}

We have examined the effect of a cylindrical object, representing a human, passing through the air curtain dividing two zones at different densities or temperatures in a corridor. The cylinder was travelling between two fixed points before and after the air curtain at different speeds $U_c$. Small-scale laboratory experiments were conducted using fresh water and salt solutions, to produce flows dynamically similar to real-scale air curtain installations.

We observed that the infiltration volume $V_i$ increases with the cylinder speed but is independent of the  density (temperature) difference  across the doorway. Moreover, the travel direction of the cylinder does not matter and we measured the same values both when the cylinder was moving from the dense fluid side to the light fluid side and vice versa. We explained this by noting that the cylinder wake is a turbulent flow field which disrupts the directed organised buoyancy-driven exchange flow. As a consequence, the infiltration flux $q_{ac}$ was also found to be independent of $\Delta\rho$.

After measuring $q_{ac}$, we calculated the effectiveness $E(D_m)$ curve of the air curtain with and without the passage of the cylinder, and observed that the effectiveness $E(D_m)$ reduces with increasing cylinder speed $U^*$ for a fixed $D_m$. In the absence of traffic, we observed that the air curtain can reduce the contaminant transport by up to about 80\% with a stable air curtain, while further increase in the curtain momentum reduces its sealing effectiveness.

At a fixed deflection modulus $D_m$, the reduction in $E$ for an increasing $U^*$ can be explained by an increasing infiltration volume $V_i$ due to the disruption of the curtain by the cylinder and the ingress of the cylinder wake. In particular, the infiltration flux $q_{ac}$ is independent of the horizontal density difference $\Delta\rho$ across the doorway, which suggested that there is no strong interaction between the cylinder wake and the gravity-driven flow. We provided a theoretical argument to explain why the reduction in $E$ is more significant at high values of $D_m$.

The cylinder disrupts the curtain and visualisations of the jet and the cylinder wake show the infiltration of the fluid carried along with the cylinder wake underneath the unestablished jet is the reason for the observed reduction in the effectiveness. With an increasing cylinder speed, the entrainment in the cylinder wake also increases due to the faster wake velocity which induces a longer time for the re-establishment of the curtain. 

{We emphasize here that the dimensionless numbers in our experiments were in the same range as for real-scale processes. We varied the deflection modulus $D_m$ from 0 to 1, which is usually the regime achieved by real air curtains \citep{FrankLinden,frank2015effects}. For a person with $l=\SI{1.7}{\meter}$ and $d=\SI{0.5}{\meter}$ moving at $U_c=\SI{1}{\meter\per\second}$, through a doorway of height $H=\SI{2}{\meter}$ and width $W=\SI{1}{\meter}$ and an air curtain discharged from $b_0=\SI{0.01}{\meter}$ and a varying outlet velocity $u_0=1 -\SI{10}{\meter\per\second}$, the $D_{m,c}$ value varies in the range 0.01 and 1.1.}

The effect of human passage on the contaminant transport is important in the design and the operation of clean rooms in chemical or pharmaceutical industries and in protecting isolated hospital rooms for infectious and immunocompromised patients from infiltration of air-borne contamination. 
Our study shows that the human or vehicular traffic reduces the effectiveness. To minimize the air curtain disruption, we suggest a slowing down of the traffic just before the air curtain and then the passage across the air curtain with a much reduced velocity. Also, a higher  safety factor on $D_m$ will result in a higher jet velocity, which will help in re-establishing the curtain faster and hence result in a lesser entrainment. 

Finally, while the geometry of the channel is directly related to a corridor in a building, as we noted in section \ref{sec:Effectiveness}, the similarity between the effectiveness as a function of the deflection modulus between these experiments and those conducted in a doorway between two spaces without confining side walls \citep{FrankLinden} suggests that the exchange is dominated by processes occurring in the immediate vicinity of the doorway, and the geometry of the spaces in either side of the doorway are of secondary importance. Thus we expect our results to hold for a doorway between two rooms or between a room and an unbounded exterior, as is often the case in practice. \\

\textbf{Acknowledgements}\\
This research has been supported by the EPSRC through grant EP/K50375/1 and Biddle BV. NKJ would gratefully acknowledge the support from PBC VATAT fellowship, Israel. We would like to thank Prof. Stuart Dalziel and Dr. Jamie Partridge for the discussions regarding the experiment. We would also like to thank D. Page-Croft for all the technical support.

\appendix

\section{Discharge coefficient}\label{app:DischCoeff}

\begin{figure}[!ht]
\centering
\includegraphics[scale=0.5]{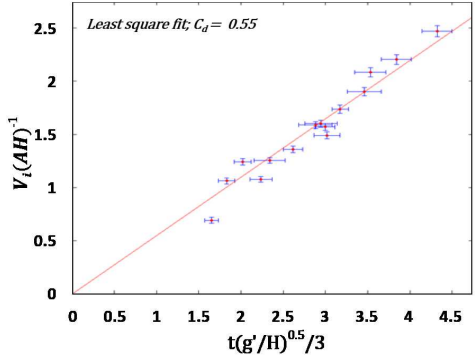}
\caption{Plot of the non-dimensionalised infiltration volume $V_i\left(AH\right)^{-1}$ as a function of the non-dimensional time $t\sqrt{g'H^{-1}}/3$ for rectangular channel. Slope of the curve is the discharge co-efficient $C_d$, which is 0.55$\pm0.03$.}
\label{Fig_Cd_a}
\end{figure}

\begin{figure}[!ht]
\centering
\includegraphics[scale=0.5]{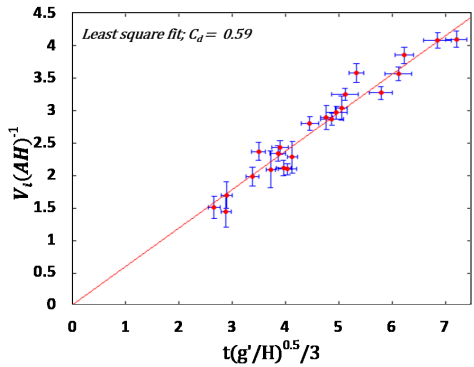}
\caption{Plot of the non-dimensional volume non-dimensionalised infiltration volume $V_i\left(AH\right)^{-1}$ as a function of the non-dimensional time $t\sqrt{g'H^{-1}}/3$ for a sharp-edged opening. Slope of the curve is the discharge co-efficient $C_d$, which is 0.59$\pm0.02$.}
\label{Fig_Cd_b}
\end{figure}

We conducted a separate series of experiments to measure the gravity-driven exchange flow through the unprotected doorway, i.e., without an operating air curtain and to determine the value for the discharge coefficient $C_d$ for (\ref{eq:orifice}).
The flux through the doorway through an unprotected doorway is proportional to the velocity scale $\sqrt{g'H}$ and the area of the opening $A$ by scaling arguments. We opted to represent the proportionality factor as $C_d/3$ (in line with a formula typically used for sharp-edged openings given in (\ref{eq:orifice})) and experimentally measured $C_d$. The proposed model with the measured value of $C_d$ describes the observed flow rate very well. {Figure \ref{Fig_Cd_a} shows the non-dimensional flow rate $V_i\left(AH\right)^{-1}$ as a function of the non-dimensional time $t\sqrt{g'H^{-1}}/3 $ as measured in our experimental setup.} The slope of the fitted linear curve is 0.55$\pm0.03$, which is the discharge co-efficient $C_d$. The discharge co-efficient for the sharp opening is also measured, which is 0.59$\pm0.02$ as shown in figure \ref{Fig_Cd_b} and is very close to previously reported value of 0.6.

\section{Supplementary materials}\label{app:SuppMat}

Video caption for supplementary video 1 (Movie1.avi): Temporal evolution of the side view of the interaction between the air curtain (blue dye) and the cylinder wake (red dye) for $U^*= 0.44$ and $D_m = 0.5$. The cylinder is moving from the dense to the light fluid side. The video is played at 4.8 times slower than the real speed.




\bibliographystyle{jfm}
\bibliography{main.bbl}

\end{document}